\documentclass[12pt]{article}
\usepackage{amscd}
\usepackage{amssymb}
\usepackage{graphicx}

\newcommand{\R}{{\mathbb R}}
\def\a{{\bf a}}
\def\k{{\bf k}}
\def\dr{{\rm d}{\bf r}}

\begin{document}

\title{\Large \bf {Ab initio structure solution by charge flipping}}
\author{G\'abor Oszl\'anyi and Andr\'as S\"ut\H o
\medskip\\
Research Institute for Solid State Physics and Optics\\
Hungarian Academy of Sciences\\
P.O.B. 49, H-1525 Budapest, Hungary\\}
\date{}
\maketitle
\thispagestyle{empty}

\begin{abstract}
\noindent
In this paper we present an extremely simple structure solution method termed \emph{charge flipping}.
It works ab initio on high resolution x-ray diffraction data in the manner of Fourier recycling.
The real space modification simply changes the sign of charge density below a threshold, while in
reciprocal space the modification is the $F_{\rm obs}$ map without any weighting. We test the algorithm
using synthetic data for a wide range of structures, analyse the solution statistics and check the quality
of reconstruction. Finally, we reconsider mathematical aspects of the algorithm in detail, showing that in
this chaotic iteration process the solution is a limit cycle and not a fixed point.
\end{abstract}

\section{Introduction}

Ab initio structure solution by x-ray diffraction is a success story of the last century.
Today the practicing crystallographer can rely on high quality data obtained on cryocooled crystals
by area detectors and synchrotron radiation. In the structure solution process freely available software
take the workload, a large part of accumulated knowledge is contained in their elaborate algorithms.
The forefront is indisputably protein crystallography where the coordinates of more than a thousand atoms
per asymmetric unit can be determined. The field of ab initio structure solution is mature, one has the
impression that there is no room for big surprises.

Nevertheless, in this paper we present an amazingly simple structure solution algorithm
 -- termed \emph{charge flipping}. This algorithm was inspired by several methods described in the literature:
phase retrieval in optics \cite{GS,Fienup,Millane}, dual space programs \texttt{SnB} and \texttt{ShelxD}
\cite{SnB,ShelxD} and recent work on iterated projections \cite{Marks,Elser}.
All these methods alternate between real and reciprocal space by the Fourier transform and do part of the job
by imposing constraints on the real space charge density. In a sense this is a return to the era before direct
methods but armed with orders of magnitude more computing power. The charge flipping algorithm described below
is probably less efficient than state of the art programs today. However, it is surprising that it works at all,
even in the small molecule world. Its extreme simplicity offers the prospect for exact mathematical formulation,
raises hopes for further improvements and may help to understand the working and limitation of other methods.

\section{On the use of data and grid}

By ab initio structure solution we mean that there is no preliminary chemical or phase information and only
a single wavelength diffraction data set is used in the process. Furthermore, anomalous scattering is not exploited,
atomic scattering factors are taken to be strictly real. In this first presentation we use synthetic data and
focus on what is feasible given ideal conditions. We assume that the diffraction data is complete
up to a given resolution, it is error free, the absolute scale and global isotropic temperature factor are known.
None of these conditions are required for the working of the algorithm, we tested the effect of Poisson noise,
the error in absolute scale and temperature factor. However, only the ideal case is discussed here to keep
the main message of the paper brief.

The real space charge density and reciprocal space structure factors are related by the Discrete Fourier Transform
which is a unitary mapping between the two spaces. For practice this is coded as variants of the high speed FFT
algorithm. If structure factors are limited by a given resolution then the charge density can be represented on a
grid without loss of information. The necessary grid spacing is $d_{\rm grid}=d_{\rm min}/2$ where
$K=2\pi/d_{\rm min}$ is the radius of resolution sphere in reciprocal space. Charge density pixels are often
calculated on a finer grid so that contour maps look better. However, this only involves a larger region
of unobserved structure factors in the calculation and does not provide more information.

The importance of finite resolution is best shown with a plot. We generated the structure factors for a typical
organic structure (example~1. of Table~1.). These are the true complex amplitudes $F(\k)$ and not just their moduli
$F_{\rm obs}(\k)$ which are used as observed data later. The resolution was set to $d_{\rm min}=0.8$\,{\AA},
structure factors outside the resolution sphere were treated as zeros. Then we calculated the
real space charge density on a 0.4\,{\AA} grid using the inverse FFT. Figure~1.a shows the sorted charge pixels.
The main characteristics is the small number of large positive values. Most pixels are concentrated around
zero, it is exactly this real space property which allows structure solution. Small negative values are naturally
present because all observed and unobserved structure factors would be needed to generate
a truly positive charge density. Behind the following algorithm the simple thought is:
positivity should be forced with care, small negative charge density may help the process of structure solution.

\section{The charge flipping algorithm}

The structure factor moduli $F_{\rm obs}(\k)$ are known for $0<|\k |\equiv k\leq K$, these are
the observed data used by the algorithm. Unobserved moduli are treated as zeros throughout the
iteration process, except $F(0)$ which is initialized to zero but later let to change freely.
The algorithm is initiated by selecting a random phase set $\{\varphi(\k)\}$ which satisfies Friedel's law
$\varphi(-\k)=-\varphi(\k)$. Structure factor amplitudes are created as $F_{\rm obs}e^{i\varphi}$
and an inverse FFT gives a real charge density $\rho({\bf r})$. This is our starting point in real space.

Then one cycle of iteration goes from real space through reciprocal to real space again
according to the following scheme:

\begin{eqnarray}\label{iter}
\begin{CD}
    \rho@>{\rm FLIP}>>g\\
    @A{\rm FFT}^{-1}AA@VV{\rm FFT}V\\
    F@<<<G
\end{CD}
\end{eqnarray}

The charge density modification $\rho({\bf r})\rightarrow g({\bf r})$ starts the current iteration cycle.
It uses a positive threshold $\delta$ for the charge pixels. The value of $\delta$
is a fraction of a typical light atom peak and is the only parameter of the algorithm.
As the height of atomic peaks depends on the grid size, data resolution and thermal parameter,
these factors also affect the choice of $\delta$. Pixels above this value are accepted unchanged on the
assumption that they belong to atomic peaks. Pixels below $\delta$ are simply multiplied by $-1$ which
is made plausible later.
In the next step temporary structure factors $G(\k)$ are calculated by an FFT. 
Then structure factors $F(\k)$ are constructed by accepting phases and replacing the moduli by $F_{\rm obs}(\k)$.
$F(0)=G(0)$ is accepted as is without fixing or limiting its value, and $F(\k)$ for $k>K$ are reset to zero.
Finally the $F(\k)$ amplitudes are inverted to obtain the new approximation of the charge density $\rho({\bf r})$.
This unconditional iteration process can continue without intervention,
the traditional $R$-factor or some other figure of merit serves only for monitoring
and not as the objective function of an optimization approach.

Symmetry is an important issue. Following positive experience \cite{Gould,Burla}, we handle all structures
in the spacegroup P1 and neglect any symmetry constraints. Accordingly, nothing fixes the origin which is
an advantage, the structure can emerge anywhere. The disadvantage is that the charge density of the whole unit cell
must be determined and not just that of the asymmetric unit. It turned out that the first factor is more important.
When we applied the symmetry constraints of a given structure and thus forced the origin to a particular pixel
of the unit cell, the success rate of the algorithm became much worse.

The algorithm is local in both spaces, modification of charge pixels and structure factors occurs only in-place. 
In real space only the charge pixels below $+\delta$ are modified which can be further divided into two parts.
Large negative values below $-\delta$ are flipped simply to force positivity. More interesting is the
$[-\delta,+\delta]$ range which is not negligible, it gives a substantial contribution to the structure factors.
Figure~1.b shows the sorted charge pixels of a typical solution. The $[-\delta,+\delta]$ range is roughly linear
which is an approximation of the target charge density. Flipping this region does not significantly change the
distribution of pixels, but at the same time sufficiently explores the phase space.
In reciprocal space the modification of structure factors corresponds to the unweighted $F_{\rm obs}$ map.
The treatment of the unobserved $F(0)$ is less standard. While its value equals the total charge and
could come from the chemical composition, we do not make use of it, keeping the algorithm ab initio
in the strict sense. $F(0)$ is initialized to be zero and is let to change freely in the iteration cycles.
In our studies this approach worked better than fixing the total charge.
Note that in this simple scheme there is no reciprocal space weighting, no tangent formula and no
use of probability. The concepts of atomicity and positivity are there but in a strange, indirect way.

During prolonged tests of the algorithm we realized that it is closely related to the solvent flipping method
of Abrahams and Leslie \cite{Abrahams1,Abrahams2} used as density modification in protein crystallography.
However, there are important differences. Solvent flipping -- as other methods of density modification --
is used for improving already existing phases and does not need atomic resolution data.
In contrast, the charge flipping algorithm of this work is used ab initio and high resolution data
is essential for its success.
Solvent flipping requires the existence of separate solvent and protein regions.
It modifies only the solvent charge density as $\rho^{\rm new}=\rho_{0}+k_{\rm flip}\cdot(\rho-\rho_{0})$,
where $\rho_{0}$ is the expected solvent level and $k_{\rm flip}$ depends on the solvent content.
In contrast, the charge flipping algorithm does not need separate real space regions,
it is applied everywhere. There is no choice of $\rho_{0}$ or $k_{\rm flip}$,
the modification is always the sign change of pixels below the threshold parameter $\delta$.
The low density region occupies the space between atoms, which is automatically found
and perpetually adjusted by the algorithm.

In the following two sections we first give several examples of ab initio structure solution
using charge flipping and then discuss mathematical aspects of the algorithm in detail.

\section{Structure solution examples}

We tested the charge flipping algorithm on more than 200 structures taken from
the Cambridge Structural Database.
For this presentation we selected ten examples in the simplest centrosymmetric and non-centrosymmetric
spacegroups each with a considerable number of atoms. The structures are listed in Table~1.

\begin{table}[h]
\caption{Example structures. Columns: CSD code and original reference, spacegroup, number of non-hydrogen atoms
and chemical formula per unit cell}
\vspace{5mm}
\centerline{
\begin{tabular}{|r|c|c|c|l|}
  \hline
  & code and ref. & spgr. & N & unit cell content \\
  \hline
  1. & \texttt{feryoq} \cite{feryoq} & P\=1 & 172 & $2\cdot C_{80}N_{1}O_{5}$\\
  2. & \texttt{rawtoy} \cite{rawtoy} & P\=1 & 216 & $2\cdot C_{88}N_{4}O_{16}$\\
  3. & \texttt{ibeyap} \cite{ibeyap} & P\=1 & 220 & $2\cdot C_{96}N_{1}O_{13}$\\
  4. & \texttt{cotgib} \cite{cotgib} & P\=1 & 244 & $4\cdot C_{53}Cu_{1}O_{5}P_{2}$\\
  5. & \texttt{sisyey} \cite{sisyey} & P\=1 & 326 & $2\cdot C_{98}Cl_{2}Mn_{12}N_{1}O_{50}$\\
  \hline
  6. & \texttt{valino} \cite{valino} & P1 & 156 & $2\cdot C_{54}N_{6}O_{18}$\\
  7. & \texttt{pawveo} \cite{pawveo} & P1 & 164 & $2\cdot C_{72}N_{4}O_{6}$\\
  8. & \texttt{gofmod} \cite{gofmod} & P1 & 188 & $2\cdot C_{77.5}N_{4}O_{12.5}$\\
  9. & \texttt{qarpuu} \cite{qarpuu} & P1 & 220 & $2\cdot C_{105}N_{4}Pd_{1}$\\
 10. & \texttt{qibbuy} \cite{qibbuy} & P1 & 240 & $1\cdot C_{181}Cl_{24}N_{6}O_{26}P_{3}$\\
  \hline
\end{tabular}
}
\end{table}

\medskip
In all cases we generated data up to 0.8\,{\AA} resolution using the coordinates and scattering
factors of non-hydrogen atoms and without adding noise. Furthermore, we assumed that the absolute scale
and the isotropic thermal parameter B are known. In practice these come from Wilson's plot.
The knowledge of the absolute scale is not a serious issue, it is simply related to the proper choice
of the $\delta$ parameter. With trial and error we can quickly find its realistic range and fine tuning
is needed only for faster convergence. In our examples ~15\% accuracy of $\delta$ is sufficient.
Using B=0 needs more consideration. Thermal vibration smears out the atomic charge density and weakens
atomicity on which the algorithm is based. Therefore, it is strongly preferred to use low temperature data.
How well the effect of B can be removed from the real data is beyond the scope of this paper but our numerical
tests show that an error of $\sim${3\,\AA}$^2$ can be tolerated.

All ten example structures were successfully solved using the charge flipping algorithm.
The solution of each structure was attempted 100 times starting with different random phase sets
and running the algorithm for a maximum of 5000 iteration cycles.
The number of iterations leading to convergence greatly varies,
only their distribution characterizes the difficulty of the problem.
Solution statistics is compiled in Table~2.

\begin{table}[h]
\caption{Solution statistics of the example structures.
Columns: $\delta$ parameter in units of $6 \cdot V_{pixel}$, success rate, mean/minimum/maximum
number of iterations}
\vspace{5mm}
\centerline{
\begin{tabular}{|r|c|c|r|r|r|}
  \hline
     & $\delta$ & success & mean & min. & max. \\
  \hline
  1. & 0.30 & 0.99 & 338 & 55 & 2005 \\
  2. & 0.28 & 1.00 & 301 & 55 & 1650 \\
  3. & 0.30 & 1.00 &  90 & 30 &  205 \\
  4. & 0.37 & 1.00 & 101 & 15 &  230 \\
  5. & 0.42 & 1.00 & 143 & 70 &  300 \\
  \hline
  6. & 0.28 & 0.95 & 1040 & 115 & 4220 \\
  7. & 0.30 & 1.00 &  106 &  35 &  345 \\
  8. & 0.30 & 1.00 &  268 &  40 & 1645 \\
  9. & 0.42 & 1.00 &  198 &  75 &  690 \\
 10. & 0.47 & 1.00 &  441 &  85 & 4115 \\
  \hline
\end{tabular}
}
\end{table}

It is informative to follow some basic quantities during the iteration.
Figure~2. shows a typical run of example~1. The three subplots are:
the total charge, the traditional $R$-factor and the phase-change.
In all three quantities a sudden decrease starts at 210 iterations and ends
after another 10 iterations. This sharp drop is an unmistakable sign of convergence
and its width is independent whether it occurs after 10 or 10000 iterations.
All curves show three different parts:
an initial transient, a long stagnation period before the convergence and an equilibrium after.
What really goes on in these periods is discussed in the next section.

Once a solution is found its quality must be evaluated. For this we locate the atoms by
$3 \times 3 \times 3$ pixel peak picking and compare their number, centroid position and
integrated weight to the original structure. The solutions are remarkably complete,
all atoms of the original structure can be found. As we work in the spacegroup P1 the
structure is always shifted relative to the original and for non-centrosymmetric structures
the solution is often the enantiomer. When we check a large number of solutions the shift
vector is uniformly distributed in the unit cell. After applying the shift and enantiomer
correction the coordinates of non-hydrogen atoms are typically within 0.1\,{\AA} from the
original structure and the integrated weight of a carbon atom scatters between 4 to 6.
This is considered very good quality reconstruction, especially without the use of
a separate refinement program.

\section{Mathematical notes on the algorithm}

Without giving a formal proof of convergence, we reconsider here some of the mathematical aspects
of the iteration method described previously.

Clearly, a prerequisite of any ab initio structure solution is that apart from translations
and point group transformations, the Fourier moduli determine a unique density.
Thus, we suppose uniqueness and mention only one obvious condition of it.
The density should not be strictly positive, otherwise any sufficiently small change in the phases
which respects Friedel's law would lead to a different, nonnegative density.

Our method assumes that the density has extended regions of zeros.
If the density
$$\rho_{\rm ideal}({\bf r})=\frac{1}{V}\sum_\k F(\k) e^{-i\k{\bf r}}$$
has a sea of zeros then the value taken by the finite sum
$$\rho({\bf r})=\frac{1}{V}\sum_{k\leq K}F(\k) e^{-i\k{\bf r}}$$
is small positive or negative near this sea of ideal zeros.
The discrete inverse Fourier transform provides a sampling of $\rho$ at the centres of pixels and not
an average over the volume of pixels. The oscillations around zero can be seen in the sampling,
and the threshold $\delta>0$ under which the sign flip is made has to be chosen in such a way that these
small oscillations fall in the interval $[-\delta,+\delta]$. Therefore, the optimal choice of $\delta$
depends on the function to be determined. We demonstrated earlier that $\delta$ can be chosen without an
{\it a priori} knowledge of $\rho$ so that the algorithm converges, and for this reason $\delta$ should not
be too small.

Given $\delta>0$, we divide $\rho$ in two parts, $\rho=\rho_{1}+\rho_{2}$ with
\begin{eqnarray}\label{rok1}
\rho_{1}({\bf r})=\left\{\begin{array}{ll}
                  \rho({\bf r})&{\rm if}\quad\rho({\bf r})\geq\delta\\
                  0&{\rm otherwise}
                  \end{array}\right.
\end{eqnarray}
and
\begin{eqnarray}\label{rok2}
\rho_{2}({\bf r})=\left\{\begin{array}{ll}
                  \rho({\bf r})&{\rm if}\quad\rho({\bf r})<\delta\\
                  0&{\rm otherwise}
                  \end{array}\right.
\end{eqnarray}

Since the input data are $F_{\rm obs}(\k)\equiv |F(\k)|$ for $0<k\leq K$,
the target function cannot be $\rho_{\rm ideal}$ but only $\rho$, including the total charge
$$F(0)=\int\rho_{\rm ideal}\,\dr=\int\rho\,\dr$$
The iteration will generate $F(0)$, although with a limited precision
because it sensitively depends on the choice of $\delta$.
If we are given $\rho_{\rm ideal}({\bf r})-\frac{1}{V}F(0)$, the charge $F(0)$ can be found by knowing
that the flat regions of $\rho_{\rm ideal}$ have to be at zero level. No similar information
about $F(0)$ can be used if we know only $\rho({\bf r})-\frac{1}{V}F(0)$.
The best we can hope for is to reproduce $\rho_{1}$ whose mere definition depends
on the positivity of $\delta$. Another fundamental reason to work with $\delta>0$
is that for any set of phases $\{\varphi(\k)\}$ and a constant $c$ large enough
\[\sum_{0<k\leq K}F_{\rm obs}(\k)e^{i(\varphi(\k)-\k{\bf r})}+c\geq 0\]
This means that any set of phases $\{\varphi(\k)\}$ is a fixed point of the iteration
if $\delta=0$ and $F(0)$ is not prescribed.\\

The scheme of iteration has been given in equation (\ref{iter}).
More precisely, we do the following:

\bigskip \noindent
{\it 0\,{\rm th} half-cycle.}

\bigskip \noindent
We choose $\varphi^{(0)}(\k)$ for (the half of) $\k$ with $0<k\leq K$ independently,
according to the uniform distribution in $[0,2\pi]$. Then
\begin{eqnarray}
F^{(0)}(\k)=\left\{\begin{array}{ll}
            F_{\rm obs}(\k)e^{i\varphi^{(0)}(\k)} & {\rm for}\ \  0<k\leq K\\
            0 & {\rm for}\ \  \k=0\ \ {\rm and}\ \  k>K.
\end{array}\right.
\end{eqnarray}
and by inverse FFT we compute $\rho^{(0)}$ determined in pixels ${\bf r}={\bf r}_j$.

\bigskip \noindent
{\it n\,{\rm th} cycle ($n\geq 1$).}

\bigskip \noindent
Given $\rho^{(n-1)}$, we divide it in two parts,
\begin{equation}\label{g}
\rho^{(n-1)}=\rho^{(n-1)}_1+\rho^{(n-1)}_2
\end{equation}
as in (\ref{rok1}) and (\ref{rok2}),
and execute the sign flip on $\rho^{(n-1)}({\bf r}_j)<\delta$ to obtain
\begin{equation}\label{ron}
g^{(n)}=\rho^{(n-1)}_1-\rho^{(n-1)}_2\ .
\end{equation}
The Fourier transform of $g^{(n)}$ provides $G^{(n)}(\k)$ for as many $\k$
as the number of pixels in the unit cell.
Then
\begin{eqnarray}\label{fnew}
F^{(n)}(\k)=\left\{\begin{array}{ll}
            F_{\rm obs}(\k)G^{(n)}(\k)/|G^{(n)}(\k)|=F_{\rm obs}(\k)
            e^{i\varphi^{(n)}(\k)} & {\rm for}\ \  0<k\leq K\\
            G^{(n)}(0) & {\rm for}\ \  \k=0\\
            0 & {\rm for}\ \  k>K.
            \end{array}\right.
\end{eqnarray}
and through inverse FFT we find the next approximation $\rho^{(n)}({\bf r})$ of the density in pixels
${\bf r}={\bf r}_j$.\\

\noindent
The real space transformation $\rho^{(n-1)}\to g^{(n)}$ is

\smallskip
\indent (i)  non-invertible, i.e. one cannot reproduce $\rho^{(n-1)}$ from $g^{(n)}$\\
\indent (ii) norm-preserving
\begin{equation}\label{norm}
\sum_jg^{(n)}({\bf r}_j)^2=\sum_j \rho^{(n-1)}({\bf r}_j)^2\
\end{equation}
\indent (iii) local in the sense that $g^{(n)}({\bf r}_j)$ depends only on $\rho^{(n-1)}({\bf r}_j)$.\\
\noindent
Together with (\ref{norm}) this implies $|g^{(n)}({\bf r}_j)|=|\rho^{(n-1)}({\bf r}_j)|$.
Its most important characteristics is, however\\
\indent (iv) sign change in a broad region of the unit cell,
$\rho^{(n-1)}({\bf r}_j)<\delta$ occurs for the majority of pixels.

\medskip
Locality in real space also means no charge displacement, implying that the position of the
density evolves freely. This has consequences on the transformation we do in reciprocal space.
Since the real space modification makes no use of symmetries, symmetry constraints on the phases
of the Fourier components are not helpful or counter-productive. As we see from equation (\ref{fnew}),
the transformation $G^{(n)}\to F^{(n)}$ in reciprocal space is also local and non-invertible but
does not preserve the norm
\begin{equation}\label{neq}
\sum_\k |F^{(n)}(\k)|^2\neq \sum_\k |G^{(n)}(\k)|^2
\end{equation}
until convergence has not reached. After convergence, instead of (\ref{neq}) equality will hold,
but not term by term. The step from $g^{(n)}$ to $G^{(n)}$ always creates nonzero Fourier components
for $k>K$, while $F^{(n)}(\k)=0$ for $k>K$. This means that
\begin{equation}
\sum_{0<k\leq K} |G^{(n)}(\k)|^2<\sum_{0<k\leq K} |F^{(n)}(\k)|^2=\sum_{0<k\leq K} F_{\rm obs}(\k)^2
\end{equation}
and therefore
\begin{equation}\label{less}
|G^{(n)}(\k)|<|F^{(n)}(\k)|=F_{\rm obs}(\k)
\end{equation}
for the largest structure factors, dominating the sum of the squares.

We emphasize that in the present algorithm convergence means reaching a limit cycle and not a fixed point,
in the sense that $|G^{(n)}(\k)|$ becomes independent of $n$, but $\varphi^{(n)}(\k)$ alternates between
two values according to the parity of $n$. In Figure 3. we plotted the evolution of $G^{(n)}(\k)$ in the
complex plane for a few strong reflections. This is more spectacular than the evolution of $F^{(n)}(\k)$
which stays on the circle of radius $F_{\rm obs}(\k)$.
Equation (\ref{g}) implies
\begin{equation}
F^{(n-1)}=F^{(n-1)}_1+F^{(n-1)}_2
\end{equation}
where $F^{(n-1)}_i$ is the Fourier transform of $\rho^{(n-1)}_i$. Thus,
\begin{equation}\label{fgg}
G^{(n)}=F^{(n-1)}_1-F^{(n-1)}_2\ .
\end{equation}
According to (\ref{fnew}), $F^{(n)}$ is a functional of $F^{(n-1)}_1-F^{(n-1)}_2$,
but not of $F^{(n-1)}$ which is not uniquely determined by $F^{(n-1)}_1-F^{(n-1)}_2$.
When convergence sets in $F^{(n)}_1(\k)$ becomes nearly independent of $n$, while
$F^{(n)}_2(\k)$ alternates between two nearly collinear vectors in the complex plane
which are nearly orthogonal to their respective $F^{(n)}_1(\k)$.
That collinearity and orthogonality are imperfect is due to (\ref{less}) and (\ref{fgg}).
As a result, we can see an even-odd alternation of $G^{(n)}(\k)$ \emph{inside} the circle
of radius $F_{\rm obs}(\k)$, as shown in Figure~4.

If the number of independent phases is $N$, the algorithm has to find one of the good phase sets
in the $N$-dimensional real space $\R^N$, where \emph{good} means reproducing a translate of
$\rho_{1}$. Any good set is represented by a point in $\R^N$, and these points form a three-
dimensional manifold $S$ having one or several connected components.
If $\rho({\bf r})\neq\rho(-{\bf r})$ then their respective translates generate
different connected sets. $\Phi=\{\varphi(\k)\}$ and $\Psi=\{\psi(\k)\}$ are in the same
component if they are connected by a space translation,
$$\psi(\k)=\varphi(\k)+\k\cdot\a$$
for some $\a\in\R^3$. Phase retrieval is done in the cube $C=[0,2\pi]^N$, and hence
parts of $S$ outside this cube have to be shifted back into it by subtracting integer
multiples of $2\pi$. Then even the connected components of $S$ fall into three dimensional
`filaments' starting and ending on the surface of the cube, and our algorithm has to converge
to a point of one of the filaments. Although $S$ is infinite, when shifted back into $C$,
the filaments do not fill densely the cube, otherwise any choice of the phases would do.
Other than a circumstantial evidence of this fact can be obtained by noting that for any
four vectors $\k_1,\ldots,\k_4$ and any $\a\in\R^3$ the numbers $\k_i\cdot\a$ are rationally
dependent, i.e. $\sum_{i=1}^4 m_i\k_i\cdot\a=0$ for suitably chosen integers $m_i$. Also,
there is a large number of shift-invariant quantities formed by the coordinates of points
of $S$, namely, if $\k'=m\k$ for some integer $m$ then
$$\varphi(\k')-m\varphi(\k)$$
is shift-invariant.

Having no \emph{a priori} information about the position of the filaments, a random initial set
of phases seems to be a good choice to start with. This is even more so, because starting with
a good phase set the iteration leaves the neighbourhood of this point and returns to another one
after convergence. Indeed, starting with the good phases implies 
$\rho^{(0)}=\rho-\frac{1}{V}\int\rho\,\dr$.
Then in real space, instead of shifting this function upwards, we start to flip the values
below $\delta$ and the iteration leads farther away from $\rho$ before it approaches again
a translate of it. That it does, is due to property (iv) of the real space transformation
which invokes a wide exploration of $C$. We note that there is no attraction along the filaments,
because the algorithm makes no preference in the position of the sample. It takes a while to reach
the basin of attraction of one of the filaments, showing that their complement in $C$ has to be a
large set of a complicated structure, similar to a `strange repellent' of a chaotic system.
However, in the neighbourhood of a solution convergence sets in like an avalanche, it is complete
after a few steps of iteration.  The iteration process is chaotic, in the usual sense of sensitive
dependence on initial conditions, which are the starting phases in our case. An obvious sign
is that the number of iteration cycles leading to convergence varies wildly with the slightest
change of a given random phase set. Chaotic behaviour comes from two sources. The topological reason
is the filamentation of the set of points to be attained in the very high dimensional cube of phases.
The algorithmic reason is that successive points to be visited in $C$ are decided by charge flipping
in real space. Although this is done in a deterministic way, during the long stagnation period
it appears very much like random jumps in the space of phases.

It is instructive to follow the evolution of the total charge $F^{(n)}(0)$ plotted in Figure~2.a.
It starts with zero, in the first step of the iteration jumps to a high positive value, which is
followed by a rapid decay to an intermediate value, a long stagnation, and a second abrupt drop to
the final number. Its evolution follows rather closely that of the $R$-factor,
\begin{equation}\label{Rfactor}
R^{(n)}=\frac{\sum_{0<k\leq K}||G^{(n)}(\k)|-F_{\rm obs}(\k)|}{\sum_{0<k\leq K}F_{\rm obs}(\k)}
\end{equation}
shown in Figure~2.b. The distance
\begin{equation}\label{dphi}
d(\Phi^{(n+2)},\Phi^{(n)})
=\frac{\sum_{0<k\leq K}F_{\rm obs}(\k)|(\varphi^{(n+2)}(\k)-\varphi^{(n)}(\k))({\rm mod}\ 2\pi)|}
{\sum_{0<k\leq K}F_{\rm obs}(\k)}
\end{equation}
exhibits a similar behaviour (Figure 2.c). We recall that the $R$-factor is used only for monitoring the
convergence, the fact that it does not tend to zero is due to (\ref{less}), and has no bearing on the
success of the iteration.

The initial overshooting of $F^{(1)}(0)$ for most of the starting phase sets
can be understood as follows. Since
$$G^{(0)}(0)=F^{(0)}(0)=0$$
and
$$F^{(0)}_1(0)=\int \rho^{(0)}_{1}\dr=-\int \rho^{(0)}_{2}\dr > 0$$
is the charge carried by the pixels in which
$\rho^{(0)}({\bf r}_j)>\delta$,
we have
$$G^{(1)}(0)=F^{(1)}(0)=2F^{(0)}_1(0)\ .$$
No similar charge doubling occurs in the subsequent steps, because typically
$$\int \rho^{(n)}_{1}\dr \gg -\int \rho^{(n)}_{2}\dr$$
for $n>0$. For the example in Figure~2. the actual limit reached by $G^{(n)}(0)$ is about 30\% lower
than the true value of $F(0)$. In principle, a smaller $\delta$ could yield the correct value,
but in practice we would not obtain convergence with a smaller $\delta$. After convergence sets in,
$\delta$ can be decreased and the iteration continued without destroying  the result. A new limit cycle
will be attained with suppressed oscillations and a higher total charge. This can go on down to
$\delta=0$, where the sea of zeros will be lost and the total charge will become too large.

\section{Conclusion}

In this paper we presented an ab initio structure solution method termed \emph{charge flipping}.
It uses high resolution data and alternates between real and reciprocal space in the manner of Fourier recycling.
The real space modification simply changes the sign of the charge density below a threshold $\delta$, which is
the only parameter of the algorithm. In reciprocal space, observed moduli are constrained using the unweighted
$F_{\rm obs}$ map while $F(0)$ corresponding to the total charge is allowed to change freely.
All structures were handled in spacegroup P1 neglecting any symmetry constraints.

We tested the algorithm using 0.8\,{\AA} resolution synthetic data for a wide range of centrosymmetric and
non-centrosymmetric structures taken from the Cambridge Structural Database. The size of our examples is
somewhere at the upper limit of small molecule structures in spacegroups P\=1 and P1. While ideal data are
not required for the working of the method, at this stage of development we needed a large pool of various
structures and we had to be sure that any difficulty of solution is not the consequence of data quality,
resolution or completeness. For the examples presented we followed single runs, analysed solution statistics
of multiple runs and checked the quality of reconstruction. Here we only emphasize that all structures were
solved with a high success rate, and all atoms were found without the use of a separate refinement program.

Finally, mathematical aspects of the iteration process were considered in detail. We clarified the most
important properties of real and reciprocal space transformations, discussed the relevance of finite
resolution and the choice of $\delta$. By following the evolution of structure factors we also showed
that the iteration process is chaotic and the solution is not a fixed point but a limit cycle.

The most important characteristics of the algorithm is its amazing simplicity. This is a big advantage
for exact mathematical treatment, in the future we shall attempt to give a formal proof of convergence.
When it comes to efficiency, such a simple method is likely to lag behind today's best programs. As long
as symmetry is not helpful, charge flipping offers only an interesting alternative in the low symmetry
spacegroups. Nevertheless, we encourage everyone to try it, it is a few lines of code plugged in an existing
program, and only a few hundred lines as a stand alone application. We shall also continue our work to check
the power of the algorithm on real data, and to make a fair comparison of success ratios to other methods.
We have well defined plans to improve the algorithm but anticipate that the original simplicity will be lost
in exchange for higher efficiency.\\

We thank Gyula Faigel, Mikl\'os Tegze and G\'abor Bortel
for useful discussions.
This research was supported by OTKA grants T043494 and T042914
and the work of G.O. was also funded by a Bolyai J\'anos Scholarship.

\begin{figure}[p]
\centerline{\includegraphics[width=8cm]{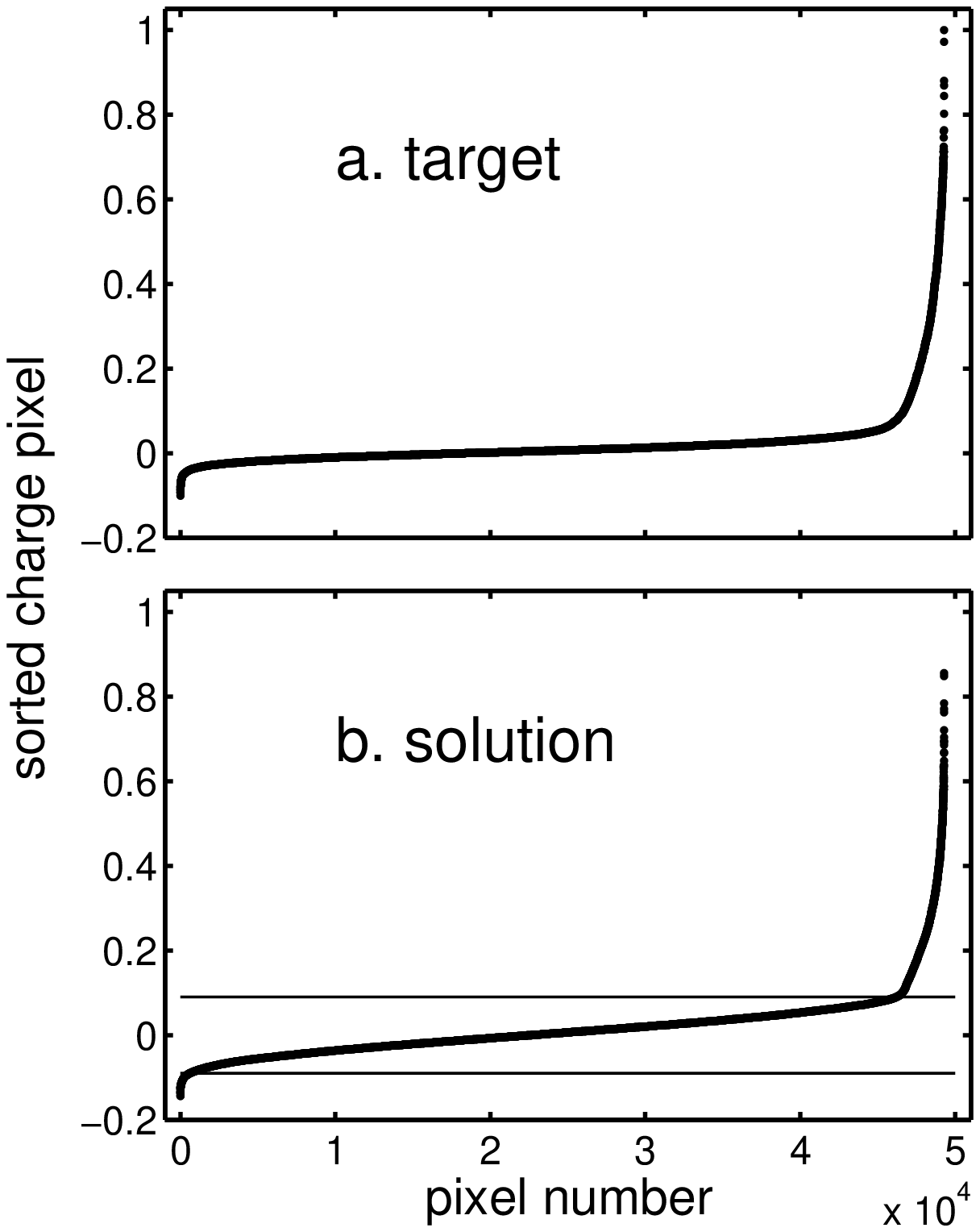}}
\caption{Charge density of a typical organic structure at 0.8\,{\AA} resolution.
a: target charge density, b: solution after convergence.
Pixel values are sorted in ascending order and are normalized to the maximum of the target.
Note that the $[-\delta,+\delta]$ range within horizontal lines is approximately linear.}
\end{figure}

\begin{figure}[p]
\centerline{\includegraphics[width=8cm]{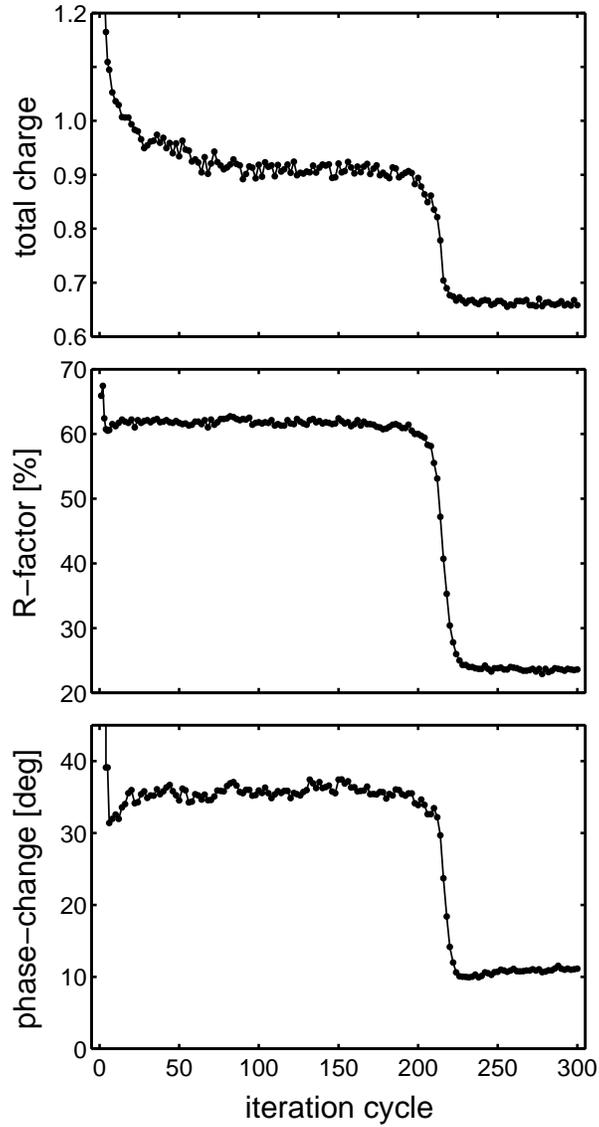}}
\caption{A typical run of the charge flipping algorithm leading to convergence.
From top to bottom: total charge, $R$-factor and phase-change as a function of the iteration cycle.
The total charge is normalized by its ideal finite resolution value.
The $R$-factor and phase-change are defined by equations (\ref{Rfactor}) and (\ref{dphi}). }
\end{figure}

\begin{figure}[p]
\centerline{\includegraphics[width=8cm]{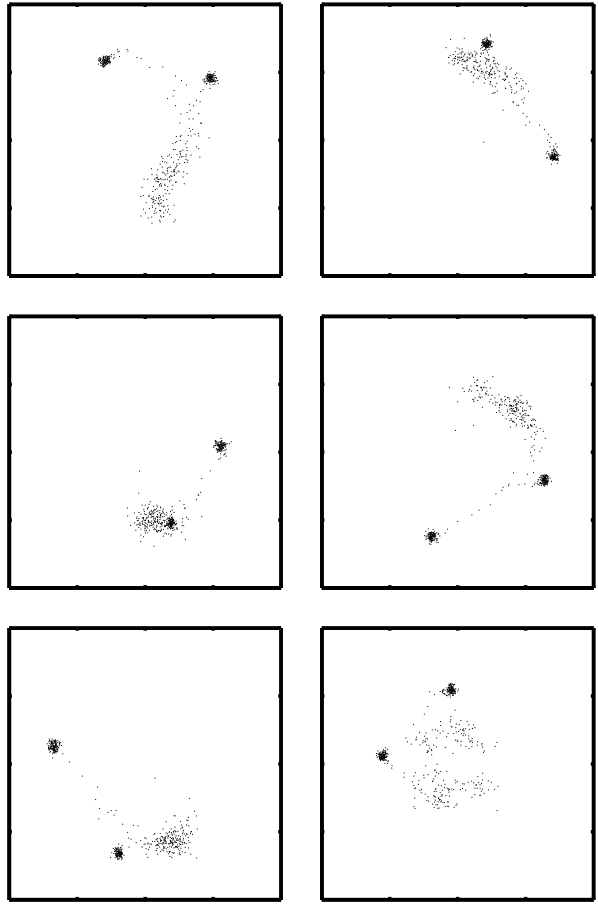}}
\caption{The evolution of the structure factor amplitude $G^{(n)}(\k)$ in the complex plane
for a few strong reflections. In each case it alternates between two values after convergence,
showing that the solution is a limit cycle.}
\end{figure}

\begin{figure}[p]
\centerline{\includegraphics[width=8cm]{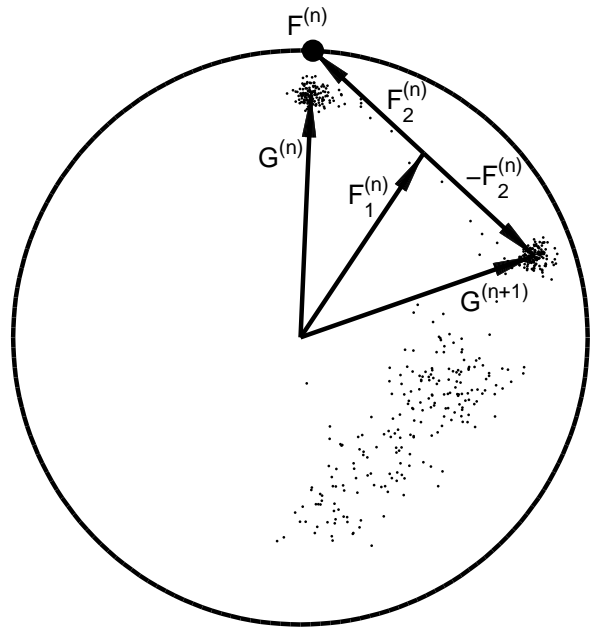}}
\caption{The evolution of a selected structure factor amplitude $G^{(n)}(\k)$ in the complex plane.
Arrows explain one cycle of iteration from $G^{(n)}$ to $G^{(n+1)}$ through
$F^{(n)}$=$F^{(n)}_1+F^{(n)}_2$. The radius of the circle is $F_{\rm obs}(\k)$ .}
\end{figure}

\end{document}